# Fully nonlinear phenomenology of the bump-on-tail (BOT) instability with drag, diffusion and Krook relaxation

Shi-Jie Zhang[1], Lei Chang[1,2*], Zhao-Ju Bo[1], Zhi-Song Qu[3], Ilya Zadiriev[4], Elena Kralkina[4], Shogo Isayama[5], Sin-Jae You[6], Zi-Chen Kan[1], Ji-Kai Sun[1], Jing-Jing Ma[1]

[1]Fundamental Plasma Physics and Innovative Applications Laboratory, School of Electrical Engineering, Chongqing University, Chongqing, 400044, China

[2]Institute of Nuclear Energy and Technology Innovation, Chongqing University, Chongqing 400044, China

[3]School of Physical and Mathematical Sciences, Nanyang Technological University, Singapore 637371, Singapore

[4]Physical Electronics Department, Faculty of Physics, Lomonosov Moscow State University, GSP-1, Leninskie Gory, Moscow, 119991, Russian Federation

[5]Department of Advanced Environmental Science and Engineering, Kyushu University, 6-1 Kasuga-Kohen, Kasuga, Fukuoka 816-8580, Japan

[6]Applied Physics lab for PLasma Engineering (APPLE), Department of Physics, Chungnam National University, Daejeon 34134, Republic of Korea

*Email: leichang@cqu.edu.cn (Lei Chang)

## Abstract

Energetic-particle-driven modes in magnetically confined fusion plasmas often exhibit nonlinear frequency sweeping ("chirping"), reflecting complex wave-particle interactions near marginal stability. While the bump-on-tail (BOT) instability within the Berk-Breizman framework has served as a canonical model for understanding such phenomena, a unified nonlinear description remains incomplete when drag, diffusion, and Krook relaxation act simultaneously. In this work, we present a comprehensive numerical investigation of the BOT instability that explicitly retains all three collision operators together with external wave damping. Using a validated characteristic-based BOT code, we systematically scan the multi-dimensional collision parameter space and construct nonlinear regime maps and bifurcation diagrams. To organize the rich dynamics, we introduce a two-level categorization that combines global wave-energy evolution (damped, steady, periodic, and chaotic states) with chirping subtypes identified from spectral morphology. We find that diffusion and Krook relaxation regularize the nonlinear dynamics and promote ordered transitions from chaotic behaviors to periodic oscillations and steady saturation as collision strength increases, with the saturation level decreasing approximately exponentially with external damping. In contrast, drag alone does not admit steady solutions and instead drives persistent or chaotic chirping through convective deformation of resonant phase-space structures. When drag is combined with diffusion or Krook relaxation, clear transition sequences emerge: increasing drag breaks hole-clump symmetry, broadens the effective resonance region, and drives systematic transitions from transient to intermittent and persistent chirping. At fixed ratios of collision operators, higher absolute collision rates require larger wave amplitudes to balance collision-driven restoration against wave-induced phase-space flattening. These results provide unified nonlinear regime maps and mechanistic phase-space interpretations for energetic-particle-driven chirping, offering a predictive framework that is directly relevant to the diagnosis and control of chirping Alfvénic activity in fusion experiments.

# 1 Introduction

In magnetically confined fusion plasmas, energetic particles (EPs) can destabilize a variety of instabilities, most notably Alfvén eigenmodes (AEs), which may in turn enhance EP transport and degrade confinement and fusion performance [1-4]. A striking feature observed in experiments is that EP-driven AEs often enter strongly nonlinear states [5-7], where rapid frequency sweeping ("chirping") emerges as a prominent and recurrent signature. Considerable progress has been made in clarifying chirping from a phase-space perspective [8], including long-range sweeping dynamics [9] and reduced descriptions based on phase-space waterbags [10], while large-scale simulations have quantified the associated convective transport in realistic configurations [11]. These advances also highlight the importance of collisional relaxation: dynamical friction can shift and split resonance lines [12], indicating that the detailed form of relaxation can fundamentally reshape resonance structures and thereby modify chirping behavior and nonlinear saturation.

To examine resonant wave-particle interaction in a controlled and analytically transparent setting, the beam-plasma instability, especially the one-dimensional bump-on-tail (BOT) model formulated within the Berk-Breizman (BB) framework, has become a canonical paradigm [13-17]. Within this model, nonlinear responses near marginal stability have been extensively classified (e.g., steady, periodic, chaotic, and explosive behaviors) and connected to phase-space hole-clump dynamics under the action of collisions and damping [7,18-25]. Such studies have established a valuable qualitative and quantitative language for interpreting chirping-like dynamics in terms of phase-space structures and their evolution. However, existing nonlinear regime maps remain incomplete for the parameter combinations most relevant to realistic relaxation. Many studies focus on restricted subsets of operators (e.g., Krook-type relaxation alone) [26,27] or explore drag and diffusion only through limited scans [28]. As a result, what is still missing is a unified nonlinear regime map when drag, diffusion, and Krook relaxation act simultaneously, precisely the situation in which different relaxation channels can compete to reshape resonant phase-space structures, alter the effective resonance condition, and change the pathway to saturation or chirping.

In this work, we present a comprehensive characterization of BOT nonlinear evolution by explicitly incorporating the competition among wave drive, external damping, and multiple collisional relaxation processes. Our simulations are carried out in a cold-background, weak warm-beam regime, where the equilibrium distribution has an approximately constant positive slope in the vicinity of the resonant velocity and the system operates in the weak-drive ordering $\gamma_L \ll \omega$, which is directly relevant to experiments [29-31]. We introduce a two-level classification that organizes nonlinear behaviors both phenomenologically and by their underlying phase-space dynamics, and we construct full multi-operator bifurcation diagrams across a multidimensional collision parameter space. Finally, we provide a phase-space interpretation of regime transitions and their sensitivity to combined relaxation, offering guidance that is directly relevant to understanding and potentially controlling EP-driven AE chirping and the associated transport in laboratory plasmas.

# 2 Model and Computation

## 2.1 Theoretical model

The BOT problem investigates nonlinear evolution of unstable electrostatic wave of frequency $\omega = \omega_{pe} = \sqrt{4\pi n_e e^2/m_e}$ in one-dimensional velocity space, in the presence of an unstable beam distribution $F(x, v, t)$ [16]. When the equilibrium distribution function $F_0$ satisfies Eq. (2.1), electrostatic wave propagating at phase velocity $\omega/k$ will be perturbed and become unstable [32].

$$\left.\frac{\partial F_0}{\partial v}\right|_{\frac{\omega}{k}} > 0. \tag{2.1}$$

In this scenario of beam-excited plasma waves, the distribution function subsequently evolves to reflect the energy transfer from the unstable distribution function to the wave itself. The system comprises three species of plasma: (i) stationary background ions, (ii) cold background particles (characterized by mass $m_e$, an equilibrium density $n_e$, and perturbed fluid velocity $V$), which provide a damping mechanism for the wave, and (iii) a small population of EPs treated kinetically. The physical essence is a resonant interaction of EPs with weakly unstable discrete modes for which the linear growth rate is much less than the mode frequency [33]. The nonlinear evolution of this kinetic system is maintained by the balance between source and relaxation processes.

Currently, nonlinear analyses of the BOT instability are primarily based on the nonlinear beam-plasma model proposed by Berk and Breizman, which incorporates source and collision effects [13-16]. This model, commonly referred to as the BB model, has been extensively developed in subsequent studies [17,20-23]. The BB model employs a quasilinear approach to solve the Vlasov-Maxwell equations, under the requirement that nonlinear modifications to high-energy particle orbits remain much smaller than the wavelength of the system's eigenmode. When collision and source terms are incorporated into the Vlasov equation, the electric field is expressed in the form $E = 1/2[\hat{E}(t)e^{i(kx-\omega t)} + c.c.]$, the kinetic equation for BOT distribution problem is given by [17]

$$\frac{\partial F}{\partial t} + v\frac{\partial F}{\partial x} - \frac{|e|E}{m}\frac{\partial F}{\partial v} = \left(\frac{dF}{dt}\right)_{\text{coll}}, \tag{2.2}$$

Maxwell's equations for the electric field is represented by

$$\left[-i\omega_{pe}\frac{\partial \hat{E}(t)}{\partial t}e^{i(kx-\omega t)} + \text{c.c}\right] + 4\pi\frac{\partial j_f}{\partial t} = 0, \tag{2.3}$$

where $\hat{E}$ is complex quantity that allows for nonlinear frequency shift, and $j_f$ represents the EPs contribution to the perturbation current driven by the wave.

Since only resonant particles contribute significantly to the nonlinear dynamics of the wave, $F$ in Eq. (2.2) can be taken as the distribution function of EPs near resonance, rather than the global distribution function [23]. The collision operator in the vicinity of resonance is expressed as

$$\frac{dF}{dt}\bigg|_{coll} = \alpha^2\left(\frac{\partial F}{\partial u} - \frac{\partial F_0}{\partial u}\right) + \nu^3\left(\frac{\partial^2 F}{\partial u^2} - \frac{\partial^2 F_0}{\partial u^2}\right) - \beta(F - F_0), \tag{2.4}$$

where $\alpha$, $\nu$ and $\beta$ are operators of drag, diffusion and Krook respectively and $u = kv - \omega$. The governing kinetic equation for the bump-on-tail problem, when all three collision operators are included, becomes

$$\frac{\partial F}{\partial t} + \left(\frac{u+\omega}{k}\right)\frac{\partial F}{\partial x} + \frac{ek}{2m}\left[\hat{E}(t)e^{i(kx-\omega t)} + c.c.\right]\frac{\partial F}{\partial u} - \nu^3\frac{\partial^2 F}{\partial u^2} \\ - \alpha^2\frac{\partial F}{\partial u} + \beta F = -\nu^3\frac{\partial^2 F_0}{\partial u^2} - \alpha^2\frac{\partial F_0}{\partial u} + \beta F_0, \tag{2.5}$$

where $F_0$ is the equilibrium distribution function in absence of any wave field, assumed to possess a constant positive velocity gradient, $\partial F_0/\partial v > 0$, which provides the linear drive rate $\gamma_L$. The constant slope assumption is valid provided that the velocity range of interest is significantly narrower than the overall width of the distribution function $F_0$ [23]. Due to the periodicity of the plasma, physical quantities can be described using Fourier analysis.

$$F = F_0(v) + f_0(v, t) + \sum_{n=1}^{N} [f_n(v, t)e^{in\xi} + c.c.], \tag{2.6}$$

Here, $\xi \equiv kx - \omega t$, $k \equiv 2\pi/\lambda$. By assuming that the temporal variation of $\hat{E}$ occurs on a time scale much slower than the wave frequency, Eq. (2.3) can be written as

$$\frac{\partial \hat{E}}{\partial t} + 4\pi e \frac{\omega}{k^2} \int f_1 du + \gamma_d \hat{E} = 0 \tag{2.7}$$

where $\gamma_d$ is wave damping. Using perturbation theory, and assuming the time scale $\tau$ of interest is sufficiently small compared to the nonlinear bounce time $\tau_B$ of particles in the wave field, the distribution function $F$ can be assumed not to deviate significantly from its equilibrium value, permitting the ordering $F_0 \gg f_1 \gg f_0, f_2$ to be taken. The perturbed distribution function is expanded as a power series in the wave amplitude $\hat{E}$.

$$\frac{\partial f_0}{\partial t} - v^3 \frac{\partial^2 f_0}{\partial u^2} - \alpha^2 \frac{\partial f_0}{\partial u} + \beta f_0 = \frac{1}{2} \frac{\partial}{\partial u}(\omega_B^2 f_1^* + c.c.), \tag{2.8}$$

$$\frac{\partial f_1}{\partial t} + iuf_1 - v^3 \frac{\partial^2 f_1}{\partial u^2} - \alpha^2 \frac{\partial f_1}{\partial u} + \beta f_1 = \frac{1}{2} \frac{\partial}{\partial u}[\omega_B^2(F_0 + f_0) + \omega_B^{2*} f_2], \tag{2.9}$$

$$\frac{\partial f_2}{\partial t} + 2iuf_2 - v^3 \frac{\partial^2 f_2}{\partial u^2} - \alpha^2 \frac{\partial f_2}{\partial u} + \beta f_2 = \frac{1}{2} \frac{\partial}{\partial u}(\omega_B^2 f_1 + \omega_B^{2*} f_3). \tag{2.10}$$

Here, $\omega_B \equiv \sqrt{|e|k\hat{E}/m_e}$, which gives the trapped particle bounce frequency. This describes the evolution of nonlinear marginally unstable mode. For a sufficiently large collision frequency this perturbative approach can be maintained indefinitely so long as $\upsilon_{eff} \gg \omega_B$. Equation (2.7) becomes

$$\frac{\partial \omega_B^2}{\partial t} - \frac{4\pi|e|^2}{m_e}\frac{\omega}{k} \int f_1 du + \gamma_d \omega_B^2 = 0. \tag{2.11}$$

It is noted that $f_1$ is initially obtained by neglecting $f_0$, providing the plasma's linear response to the wave. This $f_1$ is then used to determine $f_0$, which in turn is used to solve for an updated $f_1$, thereby yielding the first nonlinear correction to the wave amplitude. In essence, Eq. (2.8–2.10) are solved iteratively for $f_1$, which consequently leads to an iterative solution for the electric field in Eq. (2.11).

## 2.2 Numerical computation

In order to explore the full nonlinearity of the system, Eqs. (2.8–2.11) are solved numerically using a MATLAB-based computational code (also named BOT code) [17, 23, 34-38]. This code has been validated for use in BOT problem research [23]. Referring Eqs. (2.8–2.11), the model defines some normalized variables:

$$\begin{aligned} C &= \omega_B^2/\gamma_L^2, \\ G &= (2\pi \mid e \mid^2 \omega_{pe}/m\gamma_L k)F, \\ \tau &= \gamma_L t, \\ \hat{\gamma}_d &= \gamma_d/\gamma_L, \end{aligned} \tag{2.12}$$

$$\Omega = (kv - \omega_{pe})/\gamma_L,$$
$$\bar{v} = v/\gamma_L,$$
$$\gamma_L \equiv (\omega_{pe}^2 \pi / 2k^2 n_e)(\partial F_0/\partial v).$$

Using these variables, Eqs. (2.8-2.11) transform into a set of advection equations in $\Omega$:

$$\frac{\partial \mathcal{G}_n}{\partial \tau} - n \frac{\partial \mathcal{G}_n}{\partial s} + [\bar{v}^3 s^2 - i\bar{\alpha}^2 s + \bar{\beta}]\mathcal{G}_n = R_n(s,\tau) + \delta_{1,n} \frac{1}{\sqrt{2\pi}} C\delta(s), \quad (2.13)$$

$$\frac{\partial C}{\partial \tau} + \hat{\gamma}_d C = 2\sqrt{2\pi}\mathcal{G}_1(0,\tau), \quad (2.14)$$

where $\mathcal{G}_n = \frac{1}{\sqrt{2\pi}} \int_{-\infty}^{+\infty} \mathcal{G}_n \exp(-i\Omega s) d\Omega$ is the Fourier transform of the distribution function perturbation $f_n(v,t)$ with respect to velocity $v$. $n$ is the spatial harmonic number. And $C$ is the normalized wave energy (bounce frequency squared). Equations (2.13) and (2.14) constitute the final closed system solved numerically throughout this work. Since the wave growth depends on the perturbed current ($\int f_1 dv$), the integral over all $v$ is simply the value of the transform $\mathcal{G}_1(s)$ at $s=0$. $\delta_{1,n}$ is the Kronecker delta and $R_n(s,\tau)$ is the non-linearity in the equations which represents the electric force acting on the particles, and is given by

$$R_0 = \frac{is}{2}[C^*\mathcal{G}_1(s,\tau) + C\mathcal{G}_1^*(-s,\tau)], \quad (2.15)$$

$$R_n = \frac{is}{2}[C^*\mathcal{G}_{n+1}(s,\tau) + C\mathcal{G}_{n-1}(s,\tau)]. \quad (2.16)$$

The delta function $\delta(s)$ represents the constant slope of the background distribution function $F_0$. And differential collision operators are replaced by algebraic operators. Specially, by transforming variables to $s' = s + n\tau$, $\tau' = \tau$, the kinetic equations are integrated in $\tau'$ via the method of characteristics with the trapezoidal rule. and integrating in $\tau'$ by the trapezoidal rule. To avoid interpolation errors, the code forces $\Delta \tau$ to be an integer multiple of $\Delta s$ so that the shift lands exactly on a grid point.

$$\mathcal{G}_n(s, \tau + \Delta\tau) = \mathcal{G}_n(s + n\Delta\tau, \tau)\text{EXP} + \delta_{1,n}\sigma + \frac{\Delta\tau}{2}[R_n(s, \tau + \Delta\tau) + R_n(s + n\Delta\tau, \tau)\text{EXP}. \quad (2.17)$$

This approach not only substantially reduces the required resolution for $s$, but also eliminates the constraint of the Courant limit inherent to finite-differencing techniques.

$$\text{EXP} = \exp[-\frac{\bar{v}^3\Delta\tau}{3}(3s^2 + 3sn\Delta\tau + n^2\Delta\tau^2) + \frac{i\bar{\alpha}^2\Delta\tau}{2}(2s + n\Delta\tau) - \bar{\beta}\Delta\tau]. \quad (2.18)$$

Here, EXP is an integrating factor. It integrates the linear collision terms (diffusion, drag, Krook) exactly. This ensures the calculation of collisions is unconditionally stable and accurate, even for large time steps.

$$\sigma = \begin{cases} 0 & \text{if } s > 0 \\ \dfrac{C(\tau + \Delta\tau)}{2\sqrt{2\pi}} & \text{if } s = 0 \\ \dfrac{C(\tau + \Delta\tau + s)}{\sqrt{2\pi}} \exp\left(\dfrac{v^3 s^3}{3} - \dfrac{i\alpha^2 s^2}{2} + \beta s\right) & \text{if } -\Delta\tau < s < 0 \\ \dfrac{C(\tau)}{2\sqrt{2\pi}} \exp\left(\dfrac{v^3 s^3}{3} - \dfrac{i\alpha^2 s^2}{2} + \beta s\right) & \text{if } s = -\Delta\tau \end{cases} \quad (2.19)$$

The $\sigma$ term is discontinuity handling to ensure high accuracy near $s = 0$.

$$\begin{aligned} C(\tau + \Delta\tau) &= C(\tau)\exp(-\hat{\gamma}_d \Delta\tau) \\ &+ \frac{\Delta\tau}{2} 2\sqrt{2\pi}[\mathcal{G}_1(0, \tau + \Delta\tau) + \mathcal{G}_1(0, \tau)\exp(-\hat{\gamma}_d \Delta\tau)]. \end{aligned} \quad (2.20)$$

To advance the distribution function to the next time step, $\mathcal{G}_n(s, \tau + \Delta\tau)$, one would formally be required to solve a linear system of the form

$$A\,\mathcal{G}_n(s, \tau + \Delta\tau) = B\,\mathcal{G}_n(s, \tau), \quad (2.21)$$

for each value of $s$. Owing to the fact that each harmonic $\mathcal{G}_n$ interacts only with its adjacent harmonics, the matrix $A$ possesses a sparse structure. Consequently, performing a direct matrix inversion would be computationally inefficient and unnecessarily expensive. It is noting that the nonlinear contribution $R_n$ in Eq. (2.17) is proportional to $\Delta\tau$, and its magnitude further diminishes for small $s$, which is precisely the region where high numerical accuracy is most critical. This scale separation motivates the adoption of an iterative strategy to compute the updated distribution function. Accordingly, the time advancement is carried out using an iterative numerical scheme of the following form:

$$\mathcal{G}_n(s, \tau + \Delta\tau)^{(p+1)} = c + P\mathcal{G}_n(s, \tau + \Delta\tau)^{(p)}. \quad (2.22)$$

Here, $P$ denotes the iteration matrix, while $c$ is a coefficient determined by the known distribution function $\mathcal{G}_n(s, \tau)$ at the previous time level. The quantity $\mathcal{G}_n(s, \tau + \Delta\tau)^{p+1}$ represents the updated solution obtained from the $p$-th iterate $\mathcal{G}_n(s, \tau + \Delta\tau)^p$. An analogous iterative procedure is employed for advancing the wave equation. The initial iterate is naturally chosen as the solution of the corresponding linearized system, obtained by neglecting all nonlinear terms $R_n$. The parameter $F$ controls the maximum number of iterations allowed in the numerical scheme.

In this paper, all simulations are initially configured with a spatial resolution of at least 1001 grid points in $s$, maximum harmonic number $N=10$, and $\Delta\tau = \Delta s$, within a computational domain of size $s_{max} \leq 10$. The BOT code remains numerically stable given by Eq. (2.23), and when conditions of numerical non-convergence, we reduce the time step $\Delta\tau$ by increasing number of spatial grid points which decides the size of $\Delta s$. The numerical stability criterion is governed by Eq. (2.23). When the amplitude $|C|$ grows significantly, the stability conditions are no longer satisfied. To address this, we limit the maximum scale to $|s|_{\max} \sim 1/v$ (or $1/\beta$) and re-initialize the simulation. This procedure ensures both code stability and adequate resolution.

$$\Delta\tau < \frac{2}{|s|_{max}|C|} \quad (2.23)$$

## 3 Results and Discussion

Before presenting detailed parameter scans, we summarize the main physical findings of this study. First, we show that diffusion and Krook relaxation primarily act as restorative mechanisms that regularize nonlinear dynamics, promoting systematic transitions from chaotic behavior to periodic oscillations and steady saturation as collision strength increases. Second, we demonstrate that drag fundamentally differs from these mechanisms: drag alone does not admit steady solutions and instead drives persistent or chaotic chirping through convective deformation of resonant phase-space structures. Third, when multiple collision operators act simultaneously, nonlinear regime boundaries shift in a predictable manner, yielding ordered transition sequences rather than qualitatively new regimes. These results provide a unified nonlinear framework for interpreting energetic-particle-driven chirping observed in fusion experiments.

### 3.1 Categorization

As demonstrated in Sec. 2, the nonlinear evolution of the BOT instability is governed by the competition among (i) wave drive provided by resonant particles, (ii) background wave damping, and (iii) collisional relaxation processes. Because these mechanisms can dominate in different regions of parameter space, the same underlying model can evolve toward qualitatively different asymptotic behaviors. A consistent and physically motivated categorization is therefore essential, not only for describing individual simulation outcomes, but also for constructing regime maps and bifurcation diagrams in a manner that is reproducible and comparable for different collision models.

We characterize the system dynamics primarily through the normalized wave field amplitude (normalized wave energy), $\omega_B^2(t)/\gamma_L^2$, which directly quantifies the energy stored in the wave and its net exchange with resonant particles. This diagnostic is particularly suitable for parameter scans because it is robust to short transients and provides an immediate separation between decaying, saturating, and strongly time-dependent responses. To connect the wave evolution to the underlying phase-space dynamics, we further examine the distribution function in the vicinity of resonance, decomposed as $F = F_0 + f_0$, where $F_0$ is the equilibrium distribution and $f_0$ denotes the slowly varying perturbed component. Rapid transient oscillations are excluded from this analysis since they average out on long timescales and do not alter the asymptotic nonlinear state. The combined distribution $F_0 + f_0$ therefore captures the effective resonant phase-space structure-such as flattening, holes, and clumps-that ultimately determines whether wave drive can be sustained against damping and relaxation.

Based on the long-time behavior of $\omega_B^2(t)/\gamma_L^2$ after the initial transient stage, the nonlinear evolution is first classified into four primary regimes, as shown in Tab. 1, which describe the global temporal characteristics of the wave amplitude. In this work, chirping is not treated as an independent nonlinear regime, but rather as a manifestation of resonance-phase-space dynamics occurring within broader nonlinear states. This distinction is physically motivated: global wave-energy evolution reflects the balance between drive, damping, and collision-induced restoration, whereas chirping reflects localized phase-space hole–clump dynamics near resonance. Separating these two aspects allows chirping behaviour to be interpreted consistently across steady, periodic, and chaotic nonlinear regimes, resolving ambiguities present in previous single-level classifications. It is worth noting that an explosive regime has also been reported in previous studies [16]; however, since it leads to mode destabilization (unbounded growth and loss of a controlled nonlinear steady evolution), it is not considered in the present work. The above primary classification provides the backbone for identifying regime transitions and for summarizing the asymptotic response across parameter space. Nevertheless, a classification based solely on $\omega_B^2(t)/\gamma_L^2$ is not sufficient to capture the spectral manifestations of wave–particle resonance. In particular, frequency chirping can appear in multiple primary regimes-including chaotic, periodic, and even damped states-and thus requires a secondary layer of categorization that is explicitly based on frequency dynamics.

Table 1. Primary nonlinear regimes

| Regime | Criterion |
|---|---|
| Damped | The wave amplitude decays to zero within a finite time and remains zero thereafter. |
| Steady | The wave amplitude converges to a finite constant value (or asymptotically approaches a constant) without vanishing. |
| Periodic | The wave amplitude exhibits sustained oscillations with a well-defined period $T$, satisfying $\omega_B^2(t)/\gamma_L^2 = \omega_B^2(t+T)/\gamma_L^2$ |
| Chaotic | The wave amplitude remains bounded but displays irregular temporal variations, failing to satisfy any of the above criteria (i.e., no single dominant period persists over long times). |

For this purpose, we introduce a secondary categorization focusing on the characteristics of frequency chirping, based on the magnitude, persistence, and temporal organization of the frequency shift observed in the spectrogram. Within the framework of the four primary regimes, chirping behaviors are further subdivided into several types, as shown in Tab. 2. The chirping subcategories largely follow the definitions proposed by Lesur *et al.* [28], with minor consolidation adopted here for clarity and for compatibility with the broader set of behaviors observed in our parameter scans. The key distinction in the present work is that chirping is treated as a subtype spanning multiple primary regimes, rather than being restricted to a single nonlinear state. Although further distinctions could be introduced based on chirp asymmetry or fine spectral morphology, such refinements are deferred to future work to keep the present regime maps and bifurcation diagrams focused.

This two-level categorization-combining global wave-amplitude evolution with detailed spectral behavior-provides a unified and operational framework for interpreting the numerical results presented in Secs. 3.2 and 3.3. In the following sections, this framework is used to show how individual collision operators first shape the nonlinear response in isolation, and then how their combined action reshapes regime boundaries and drives transitions among categorized behaviors.

Table 2. Chirping subcategories

| Subtype | Criterion |
|---|---|
| Damped chirping | Frequency sweeping occurs transiently and disappears permanently after a finite time. |
| Persistent chirping | The spectrogram is rarely quiescent, with a long-lived hole or clump dominating the spectrum and inducing a sustained frequency shift $\delta\omega_{\max}$. Depending on whether $\delta\omega_{\max}$ oscillates or remains constant, the structure corresponds to oscillatory or steady holes (or clumps). |
| Periodic chirping | Frequency sweeping repeats regularly with a well-defined period $T$, satisfying $\delta\omega(t) = \delta\omega(t+T)$. |
| Intermittent chirping | The spectrogram alternates between quiescent intervals and active chirping phases, each containing one or more chirping branches. |
| Chaotic chirping | The spectrogram is rarely quiescent, but the frequency shifts remain small, with no long-lived hole or clump structures, or the behavior does not conform to any of the above categories. |

## 3.2 Nonlinear behavior with individual collision effect

Having established the classification framework for nonlinear regimes and chirping behaviors in Sec. 3.1, we now apply this scheme to quantify how individual collisional relaxation mechanisms modify the nonlinear evolution of the BOT instability. This step is necessary because the three operators appearing in Eq. (2.4), i.e., diffusion, Krook, and drag, represent distinct physical actions in velocity space (gradient smoothing, direct relaxation, and convective transport, respectively). By isolating each operator while suppressing the others, we can identify its characteristic impact on (i) wave amplitude evolution, (ii) frequency dynamics, and (iii) the formation and survival of phase-space structures, without ambiguity from competing effects. Moreover, near the instability threshold ($\gamma_L - \gamma_d \ll \gamma_L$), frequency chirping and hole–clump structures may arise even in the collisionless limit, indicating that resonant wave drive alone can sustain nonlinear dynamics for sufficiently long times. Once a collision operator is introduced, however, the balance between wave drive, damping, and relaxation is altered in a controlled way, and the system may transition among the primary regimes and chirping subtypes defined in Sec. 3.1. Establishing these single-operator "response templates" is therefore essential: they provide the baseline for interpreting the richer multi-operator phenomenology discussed later in Sec. 3.3. Accordingly, in the following subsections we examine diffusion, Krook, and drag collisions one at a time. For each case, we first present the wave-amplitude evolution and identify the corresponding primary regime; we then analyze the frequency dynamics via spectrograms and classify any chirping as a secondary subtype; finally, we connect these observations to the evolution of the distribution function $F = F_0 + f_0$, highlighting the phase-space mechanisms responsible for the observed transitions.

### 3.2.1 Effect of diffusion collision

We first consider the effect of velocity-space diffusion, which acts to smooth fine-scale structures in the distribution function and thereby suppress strong nonlinear phase-space gradients. From a physical standpoint, diffusion tends to erase narrow holes/clumps and promotes flattening over a broader velocity interval, which can regularize the nonlinear response and favor saturation. Consistent with this expectation, Fig. 1 shows the wave evolution near the threshold for three representative diffusion strengths (with the same background damping condition as indicated in the figure caption). For a relatively weak diffusion strength, $\nu/(\gamma_L - \gamma_d) = 1.6$, the wave exhibits bounded but irregular amplitude fluctuations, corresponding to the chaotic regime. Increasing diffusion to $\nu/(\gamma_L - \gamma_d) = 2.07$ regularizes the dynamics and produces sustained oscillations with a well-defined period, i.e., a transition to the periodic regime. With further increase to $\nu/(\gamma_L - \gamma_d) = 2.6$, the oscillations are suppressed and the system approaches a finite constant amplitude, yielding a steady-state solution. These transitions map directly onto the primary regimes defined in Sec. 3.1, illustrating that diffusion provides an efficient pathway from irregular nonlinear dynamics to regular saturation.

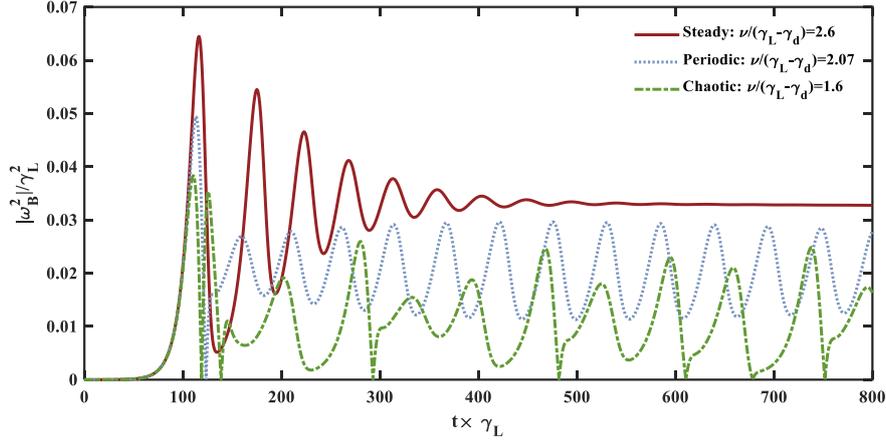

Figure 1. Temporal evolution of the normalized wave energy $C(\tau) \equiv |\omega_B^2/\gamma_L^2|$ as a function of normalized time $\tau = \gamma_L \times t$ for three diffusion strengths $\nu/(\gamma_L - \gamma_d) = 1.6, 2.07,$ and $2.6$ at the near-threshold condition $\gamma_d/\gamma_L = 0.9$. The three cases exemplify, respectively, chaotic, periodic, and steady-state responses according to the categorization in Sec. 3.1.

To connect these amplitude-level observations with phase-space evolution, Fig. 2 shows the corresponding behavior of the slowly varying perturbed distribution function $f_0$. As diffusion increases, the flattened region around the resonance broadens, which reduces the effective drive available to sustain strong oscillatory or irregular dynamics. In addition, because $f_0$ feeds into the evolution of the higher harmonics through Eqs. (2.7-2.10), changes in $f_0$ directly modify the nonlinear correction to the wave field. In this sense, enhanced diffusion does not merely "damp the wave"; rather, it modifies the self-consistent wave-particle coupling by limiting the buildup and persistence of phase-space gradients, promoting saturation and stabilizing the long-time response.

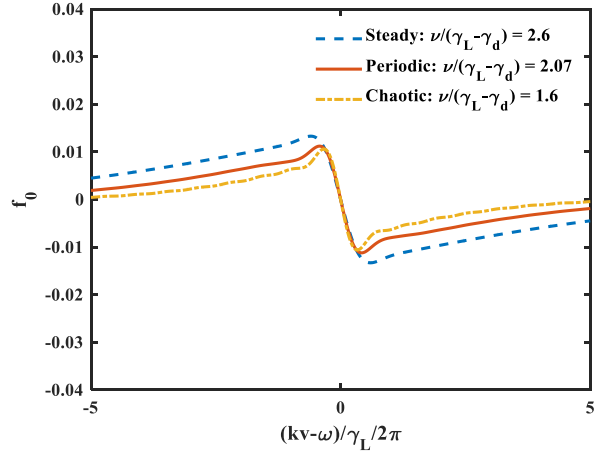

Figure 2. Profiles of the slowly varying perturbed distribution $f_0$ in the resonant region for the three diffusion cases shown in Fig. 1, plotted versus the normalized detuning $\Omega/2\pi$ with $\Omega \equiv (kv - \omega)/\gamma_L$ (resonance at $\Omega = 0$). The dependence of the resonant slope and the width of the flattened region on $\nu$ illustrates how diffusion regulates phase-space structure and thus nonlinear regime.

Although chaotic regimes can in principle support chirping, sufficiently strong diffusion suppresses long-lived hole-clump structures and therefore prevents sustained frequency sweeping. When diffusion is reduced to smaller values, transient frequency shifts can still occur, but they tend to terminate as the mode decays, corresponding to the damped chirping subtype introduced in Sec. 3.1. Because such transient chirping embedded in an ultimately decaying response has received less attention than persistent chirping regimes, we analyze this behavior in detail by tracking the instability evolution and the distribution function at several characteristic time instances. Figure 3 illustrates a representative case at $\nu/(\gamma_L - \gamma_d) = 1$ (with the damping ratio specified in the figure). The electric-field amplitude is bounded and irregular at early times (Fig. 3(a)), and a burst of activity around $t \times \gamma_L \approx 200$ is accompanied by upward and downward frequency shifts in spectrogram (Fig. 3(b)). However, field oscillations and frequency sweeping vanish by $t \times \gamma_L \approx 530$, indicating that the system ultimately enters a self-decaying state despite the presence of transient chirping.

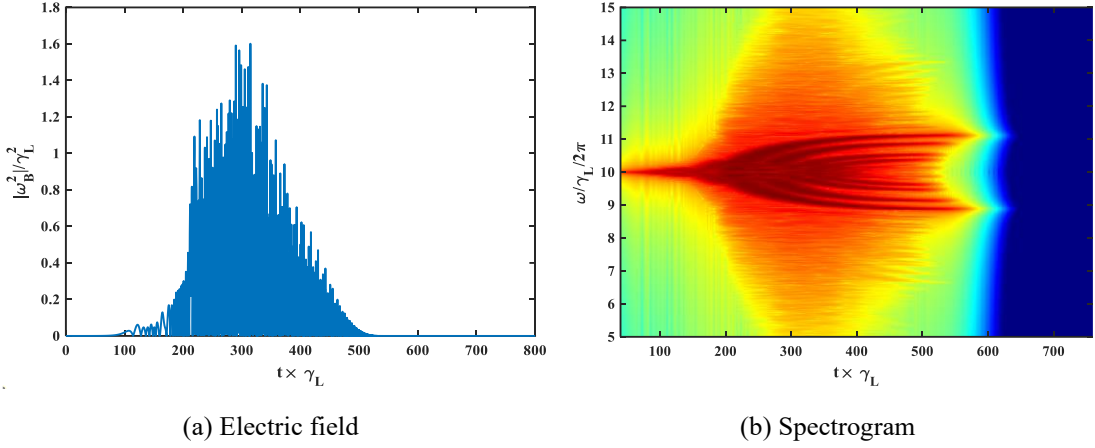

(a) Electric field    (b) Spectrogram

Figure 3. Example of a diffusion-dominated case exhibiting transient chirping ("damped chirping") at $\nu/(\gamma_L - \gamma_d) = 1$ and $\gamma_d/\gamma_L = 0.9$: (a) time trace of $C(\tau)$; (b) spectrogram (short-time Fourier transform) of the wave field intensity, plotted in normalized frequency $\omega/\gamma_L/2\pi$. The frequency sweeping terminates as the wave amplitude decays to zero, consistent with the damped primary regime defined in Sec. 3.1.

The evolution of the full distribution function in Fig. 4 further clarifies the mechanism behind this damped chirping. At $t \times \gamma_L \approx 0$, the distribution coincides with the equilibrium $F_0$. During the interval $t \times \gamma_L \simeq 0\text{-}300$, a hole-clump pair forms and propagates away from the resonance toward both sides, producing simultaneous upward and downward frequency sweeping as observed in Fig. 3(b). As the pair evolves, phase-space flattening develops near resonance, which reduces the local effective growth rate of wave and weakens the drive sustaining spatial perturbations. Consequently, once background damping becomes dominant, the wave amplitude decreases over $t \times \gamma_L \simeq 300\text{-}530$ and chirping terminates. Comparing distribution functions at $t \times \gamma_L \approx 530$ and $t \times \gamma_L \approx 800$ shows that diffusion continues to smooth the distribution on both sides of resonance; meanwhile, the slope at the resonance remains similar between these times but is already significantly flattened compared to the initial equilibrium. The resulting slope is therefore insufficient to re-excite the BOT instability, explaining why the mode does not recover after the chirping burst.

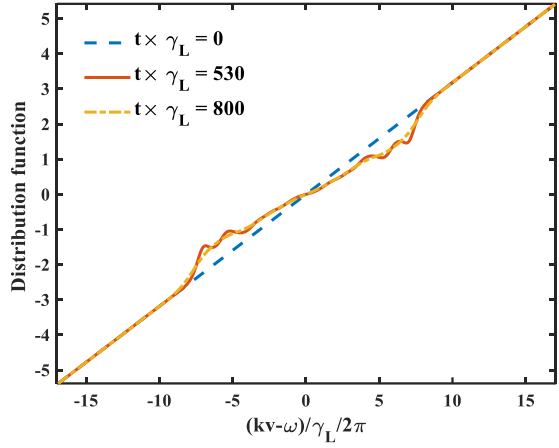

Figure 4. Time evolution of the (normalized) distribution function $F = F_0 + f_0$ for the case in Fig. 3, shown at $t \times \gamma_L = 0$, $t \times \gamma_L \approx 530$, and $t \times \gamma_L \approx 800$ as a function of $\Omega/2\pi$ ($\Omega \equiv (kv - \omega)/\gamma_L$). The transient hole-clump formation and subsequent resonance-region flattening explain the observed burst of chirping and its eventual cessation.

Up to this point, the discussion has focused on diffusion-driven regime transitions at fixed damping conditions near threshold. However, according to Eq. (2.11), the temporal evolution of the electric field depends not only on the resonant response (through $f_1$) but also explicitly on the background dissipation $\gamma_d$. It is therefore instructive to separate the roles of diffusion and damping by examining how the nonlinear saturation amplitude varies across the damping parameter space while remaining within the same primary regime. Figure 5 shows the wave evolution for different ratios $\gamma_d/\gamma_L$ while keeping $\nu/(\gamma_L - \gamma_d) = 2$ to ensure all cases remain in the steady-state regime. As $\gamma_d/\gamma_L$ decreases, the saturated wave amplitude increases, indicating that weaker background damping allows the wave to extract more energy from resonant particles before equilibrium is reached. This trend is consistent with the corresponding distributions shown in Fig. 6: larger wave amplitude implies that a broader population of resonant particles participates in wave-particle interaction, resulting in a wider flattened region. The flattening reduces the available drive, and the system ultimately reaches a saturated equilibrium through the balance between wave-induced flattening and combined damping/relaxation processes.

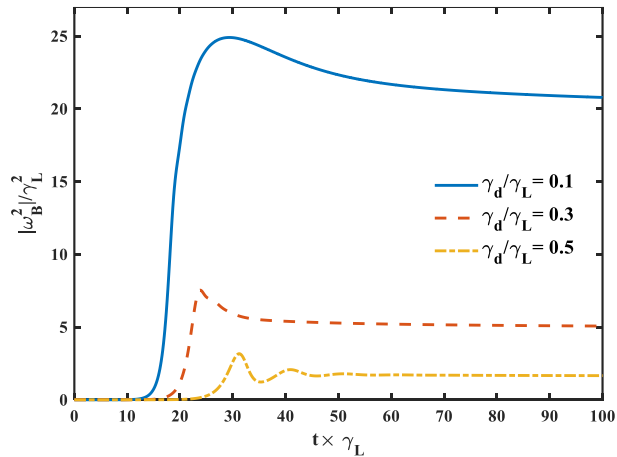

Figure 5. Effect of background damping on diffusion-dominated steady-state saturation: time traces of $C(\tau)$ for $\gamma_d/\gamma_L = 0.1$, 0.3, and 0.5 with diffusion fixed at $\nu/(\gamma_L - \gamma_d) = 2$, chosen such that all cases remain in the steady-state regime.

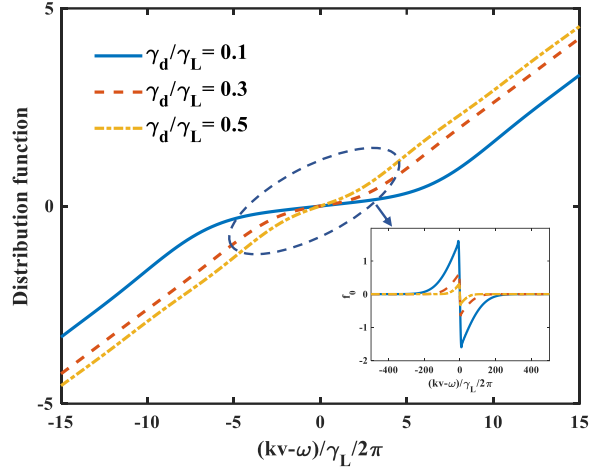

Figure 6. Late-time (steady-state) distribution functions $F(\Omega) = F_0 + f_0$ corresponding to Fig. 5 for different $\gamma_d/\gamma_L$, plotted versus $\Omega/2\pi$. The inset shows the corresponding $f_0$, highlighting the resonance-region modification (flattening) that balances drive and damping. Both $F$ and $f_0$ are normalized as described in the text.

To quantify this dependence more systematically, Fig. 7 summarizes the steady-state amplitude variation as $\gamma_d/\gamma_L$ is scanned from 0.1 to 0.99 at fixed $\nu/(\gamma_L - \gamma_d) = 2$. The wave amplitude exhibits a clear negative exponential dependence on $\gamma_d$, consistent with the damping term in Eq. (2.11). This damping-scan result under diffusion-only conditions provides a useful baseline for the next subsection, where we repeat the analysis for Krook collisions and compare how different restoration mechanisms modify both the magnitude of the perturbation and the route to saturation.

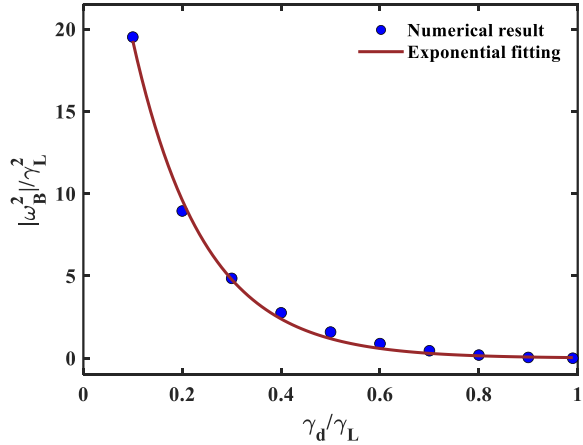

Figure 7. Steady-state saturation level $C_{\text{sat}}$ as a function of $\gamma_d/\gamma_L$ for the diffusion-only model at fixed $\nu/(\gamma_L - \gamma_d) = 2$.

### 3.2.2 Effect of Krook collision

We next consider the case where only the Krook collision operator is retained (i.e., $\nu = 0$ and $\alpha = 0$), so that relaxation acts directly on the distribution function toward its equilibrium form.

Similar to diffusion, Krook collisions suppress nonlinear structures and can promote saturation; however, because the Krook operator acts on $F - F_0$ itself rather than on its velocity gradient, the resulting phase-space evolution and saturation pathway can differ quantitatively.

Near the threshold, the qualitative regime sequence under Krook collisions resembles that observed for diffusion in Fig. 1, as shown in Fig. 8. Under the same background damping and comparable normalized operator strength, the key difference is that the Krook collision produces a larger perturbation level, which can be verified by comparing the distribution functions in Fig. 9 with those in Fig. 2. Physically, this reflects the fact that diffusion efficiently removes fine-scale gradients and tends to suppress localized structures, whereas the Krook operator allows nonlinear structures to develop but continuously relaxes the overall perturbation toward equilibrium. As a result, the wave can reach a relatively higher amplitude before the balance between drive, damping, and relaxation is achieved [35].

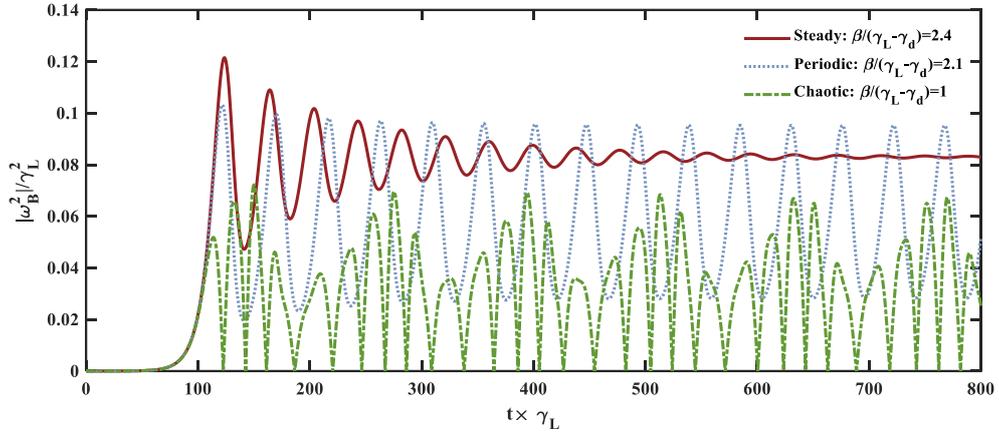

Figure 8. Temporal evolution of $C(\tau)$ for three Krook relaxation rates $\beta/(\gamma_L - \gamma_d) = 1.0, 2.1$, and 2.4 at $\gamma_d/\gamma_L = 0.9$. The cases represent chaotic, periodic, and steady-state nonlinear responses, illustrating how Krook relaxation alters regime accessibility near threshold.

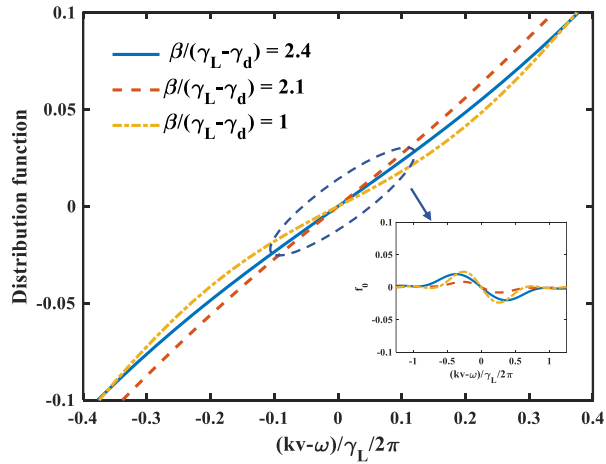

Figure 9. Resonant-region distribution-function response under Krook relaxation for the cases in Fig. 8, plotted versus $\Omega/2\pi$. The inset displays the corresponding perturbed component $f_0$.

To clarify how Krook relaxation shapes the approach to saturation, Fig. 10 illustrates two representative steady-state behaviors: (a) an underdamped case, where the wave undergoes damped oscillations before converging to a constant amplitude, and (b) an overdamped (or critically damped) case, where the wave relaxes monotonically to the saturated level without oscillation. By scanning $\gamma_d/\gamma_L$ while keeping $\beta/(\gamma_L - \gamma_d) = 0.5$, one observes that stronger damping favors overdamped relaxation, whereas weaker damping allows oscillatory transients to persist for longer times. Importantly, in both cases the system ultimately falls into the same steady-state primary regime of Sec. 3.1; the difference lies in the transient pathway by which the asymptotic equilibrium is reached.

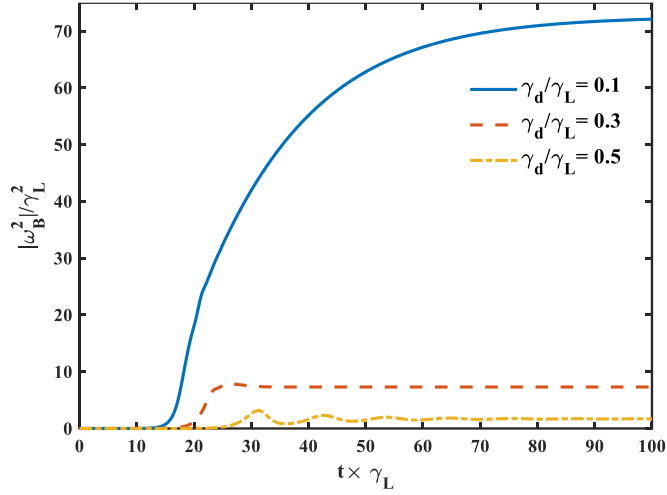

Figure 10. Damping dependence of Krook-dominated steady-state saturation: time traces of $C(\tau)$ for $\gamma_d/\gamma_L = 0.1$, 0.3, and 0.5 at fixed $\beta/(\gamma_L - \gamma_d) = 0.5$. All cases converge to steady saturation, with lower damping leading to higher $C_{\text{sat}}$.

The corresponding distribution-function evolution shown in Fig. 11 provides the phase-space interpretation of these behaviors. Similar to the diffusion-only case (Fig. 6), the drive from resonant particles initially dominates background damping, and phase mixing together with the formation of phase-space islands leads to flattening near resonance. Under Krook relaxation, however, the perturbation is continuously pulled back toward the equilibrium distribution, which modifies the width and depth of the flattened region and thereby affects the effective wave drive sustaining the mode. This interplay explains why the wave may display damped oscillations (underdamped approach) before stabilizing, even though the final state is steady.

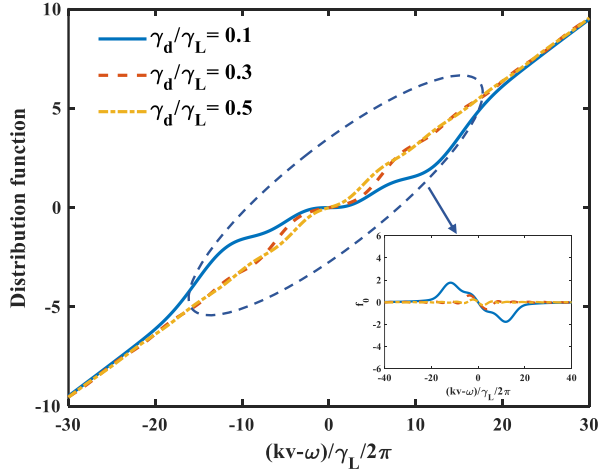

Figure 11. Late-time distribution functions $F = F_0 + f_0$ corresponding to Fig. 10, plotted versus $\Omega/2\pi$ The inset shows $f_0$, illustrating how the resonance-region flattening broadens as the saturation amplitude increases, consistent with stronger wave–particle interaction at lower $\gamma_d/\gamma_L$.

Finally, Fig. 12 demonstrates that under the single Krook collision model the steady-state wave amplitude also exhibits a negative exponential dependence across the damping parameter space. Taken together, Secs. 3.2.1 and 3.2.2 show that both diffusion and Krook relaxation can produce steady-state saturation, but they do so through different phase-space restoration pathways and with different perturbation magnitudes. This distinction becomes crucial when convective effects are present, because drag tends to deform resonance structures rather than simply suppress them. We therefore turn next to the drag-only case, which exhibits a qualitatively different nonlinear phenomenology.

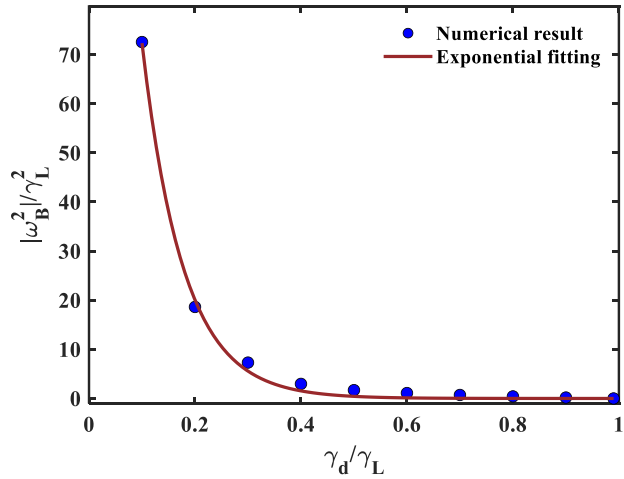

Figure 12. Steady-state saturation level $C_{\text{sat}}$ versus $\gamma_d/\gamma_L$ for the Krook-only model at fixed $\beta/(\gamma_L - \gamma_d) = 0.5$.

### 3.2.3 Effect of drag collision

Finally, we consider the case of single drag collision, which differs fundamentally from diffusion and Krook operators. Drag acts convectively in velocity space, effectively shifting resonant structures and breaking time-reversal symmetry; as a consequence, it introduces "memory" into the phase-space evolution and can continuously transport resonant particles. This convective nature is typically destabilizing in near-threshold systems and is expected to favor strong time dependence and chirping rather than steady saturation.

Consistent with this physical picture, no steady-state solution is observed within the drag parameter space, regardless of how small the normalized drag operator $\alpha/(\gamma_L - \gamma_d)$ is chosen. Instead, the system exhibits either chaotic solutions (as exemplified in Fig. 13) or explosive solutions. In these cases, the wave amplitude can reach higher levels than in diffusion-only or Krook-only systems, and under certain conditions it undergoes rapid nonlinear growth rather than approaching a stable saturated state.

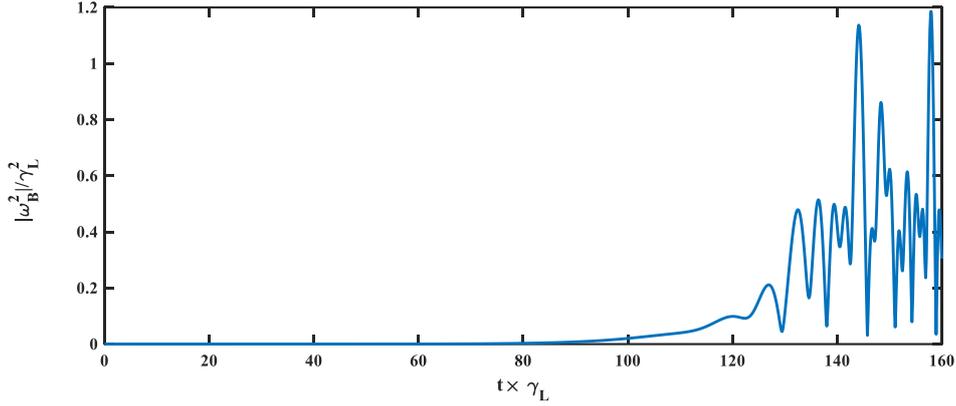

Figure 13. Drag-only nonlinear evolution of the normalized wave energy $C(\tau)$ for $\alpha/(\gamma_L - \gamma_d) = 1.2$ (other parameters are same with Fig. 8).

The spectral and phase-space signatures of this drag-dominated response are shown in Fig. 14. In both the chaotic and explosive regimes, intense frequency chirping is observed in the spectrogram (Fig. 14(b)), indicating persistent and strong wave–particle resonance dynamics. Simultaneously, the perturbation of the distribution function and the effective resonance region continuously shift in velocity space (Fig. 14(a)), and the local gradient at resonance may even reverse sign. Such behavior implies that the assumptions underlying marginal stability theory are no longer adequate, signaling the onset of a strongly nonlinear regime [35]. In the terminology of Sec. 3.1, these responses naturally fall into chirping-dominated subtypes (typically chaotic chirping or persistent chirping), and they highlight that drag does not merely "relax" the distribution but actively reshapes resonance structures through convective transport.

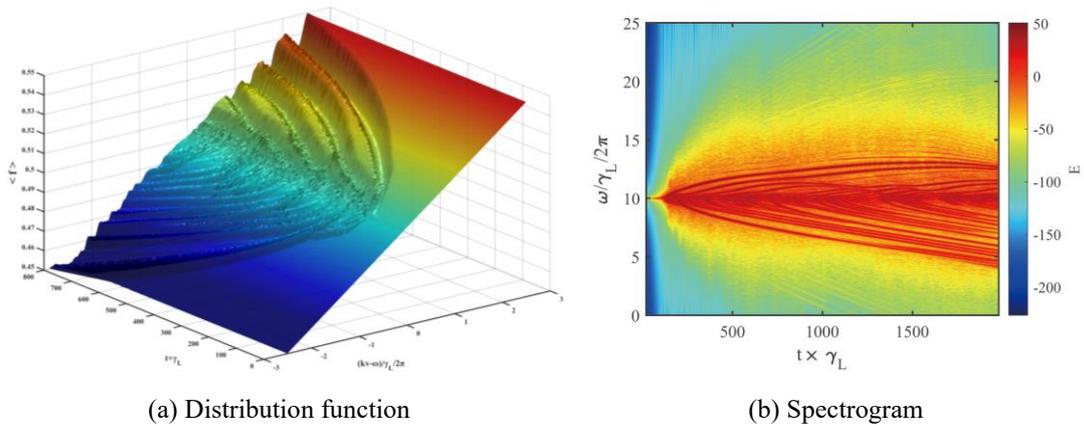

(a) Distribution function        (b) Spectrogram

Figure 14. Nonlinear phase-space dynamics and spectral signature for the drag-only case in Fig. 13: (a) three-dimensional evolution of the distribution function $F(\Omega, \tau)$; (b) corresponding spectrogram showing strong, persistent chirping.

This qualitative contrast between drag and the two restorative operators (diffusion and Krook) sets the need and motivation for Sec. 3.3: when drag is combined with diffusion or Krook relaxation, the resulting nonlinear behavior is determined by the competition between convective deformation (which sustains chirping) and restorative smoothing/relaxation (which suppresses coherent phase-space structures). The multi-operator results therefore cannot be interpreted reliably without the single-operator baselines established in the present section.

### 3.3 Nonlinear behavior with multiple collision effects

To maintain continuity with the categorization framework of Sec. 3.1, we shall interpret multi-operator results using the same two-level classification. Methodologically, we focus first on near-threshold conditions where chirping and hole-clump formation are most prominent, because this is the parameter region in which the balance between drive, damping, and relaxation is particularly delicate. We then complement the analysis by examining how saturation amplitude varies when the overall strength of collisions is increased while keeping the ratio between operators fixed, which provides a quantitative link to the restoration-versus-flattening balance emphasized in Eq. (2.11).

### 3.3.1 Effects of diffusion and drag collisions

We first investigate the combined influence of diffusion and drag, because these two operators represent opposing tendencies: diffusion suppresses fine phase-space structure, whereas drag promotes convective transport and tends to sustain chirping. Our goal here is to isolate how adding drag modifies the chirping dynamics that, under diffusion alone, may be transient or suppressed. To achieve this in a controlled manner, we adopt the parameter conditions corresponding to the chirping case discussed earlier (i.e., the same baseline conditions as in Fig. 3(b)), and we keep the system initially within a regime that is capable of exhibiting chirping. In practice, this is done by maintaining the diffusion operator at $\nu/(\gamma_L - \gamma_d) = 1$ while varying the drag operator, so that the role of drag can be identified unambiguously. Figure 15 shows the three-dimensional evolution of the distribution function near the threshold $\gamma_d/\gamma_L = 0.9$ for three representative drag strengths. When only small drag term is introduced, $\alpha/(\gamma_L - \gamma_d) = 0.5$, the distribution develops a transient hole-clump structure (Fig. 15(a)), resembling the early-stage behavior seen in diffusion-dominated cases. However, as drag operator increases to $\alpha/(\gamma_L - \gamma_d) = 1$ (i.e., $\alpha/\nu = 1$ at fixed $\nu/(\gamma_L - \gamma_d) = 1$), the perturbation loses its symmetry (Fig. 15(b)). This asymmetry is a key signature of drag: drag suppresses clump formation while favoring hole growth, thereby biasing the phase-space evolution toward hole-dominated structures rather than symmetric hole-clump pairs. As a consequence, multiple discrete holes appear and stack in velocity space, progressively flattening the distribution around resonance. This flattening reduces the net drive from resonant particles, which directly impacts both the wave amplitude and the organization of chirping in the spectrum.

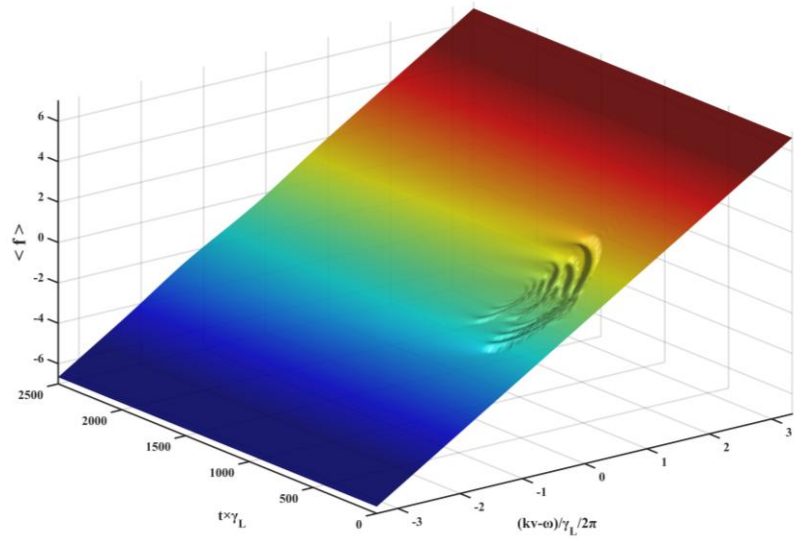

(a)

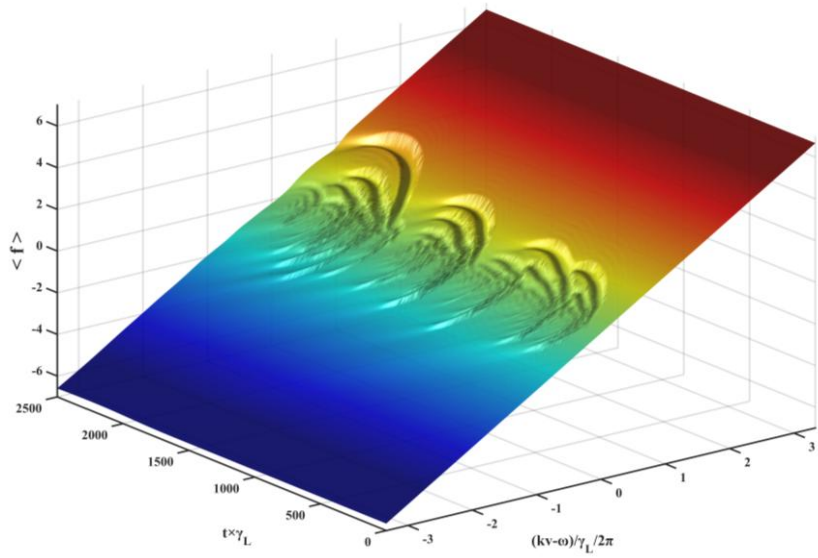

(b)

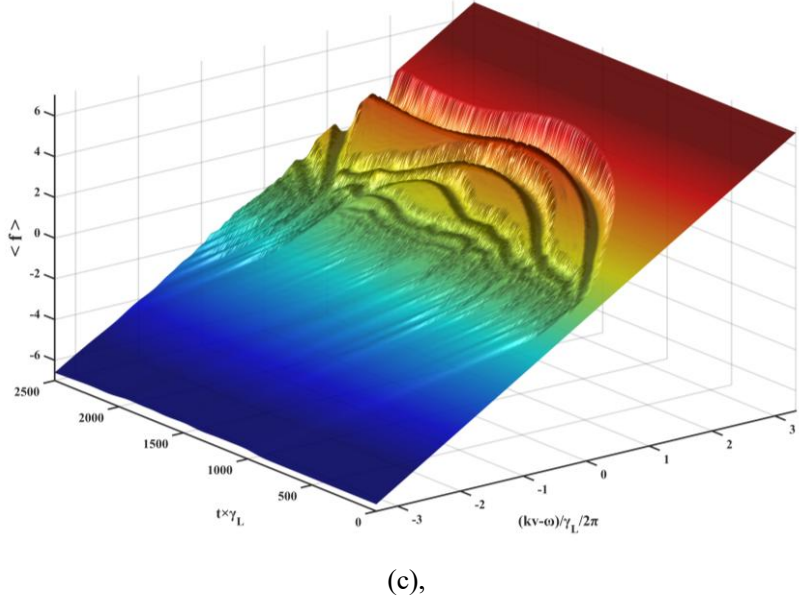

(c),

Figure 15. Three-dimensional evolution of the distribution function under combined diffusion and drag collisions at $\gamma_d/\gamma_L = 0.9$ with diffusion fixed at $\nu/(\gamma_L - \gamma_d) = 1$: (a) $\alpha/(\gamma_L - \gamma_d) = 0.5$; (b) $\alpha/(\gamma_L - \gamma_d) = 1.0$; (c) $\alpha/(\gamma_L - \gamma_d) = 1.5$.

The above phase-space evolution is reflected more clearly in the wave spectrum. Figure 16(a) shows a sudden termination of chirping that occurs at the same time as in Fig. 15(a). Importantly, this cessation is not merely a spectral artifact: it coincides with the decay of the electric-field amplitude to zero, indicating that the system is in a damped primary regime with damped chirping as a transient subtype. The mechanism is consistent with the near-threshold balance discussed in Sec. 3.2.1: under strong background dissipation, if the effective wave drive is weakened (here by diffusion-assisted flattening and insufficient drag-induced broadening), the field cannot be maintained at a finite level. When drag is increased, however, the resonance range is broadened and more energy can be transferred from energetic particles to the wave, leading to more sustained chirping activity and a characteristic hook-like chirp pattern as shown in Fig. 16(b). Notably, such hook-like patterns resemble chirping structures reported in experiments [39], supporting the physical relevance of the combined-operator dynamics. As the drag operator is further increased to $\alpha/\nu = 1.5$, the distribution evolves toward a stable hole-dominated structure (Fig. 15(c)), and the upward frequency sweeping saturates (Fig. 16(c)). In the language of Sec. 3.1, this progression corresponds to a systematic transition from damped chirping (chirp terminates as the field decays), to chaotic chirping (chirp persists with irregular organization), and finally to persistent chirping (long-lived hole dominance and sustained frequency shift). This sequence highlights a central multi-operator principle: diffusion controls the survivability of phase-space structures, while drag controls their convective persistence and asymmetry; together, they determine whether chirping is transient, intermittent, or sustained.

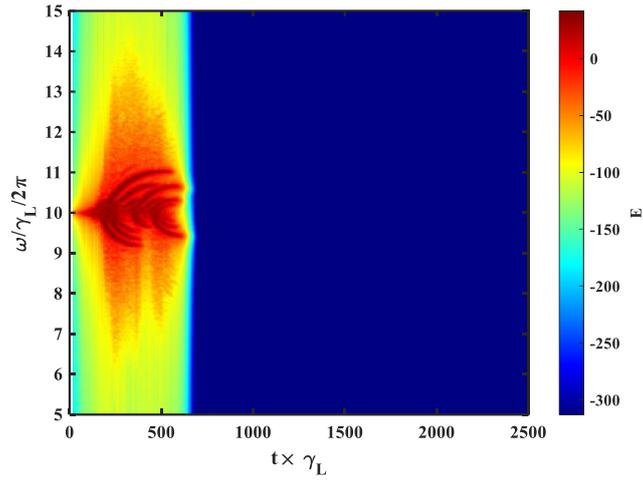

(a)

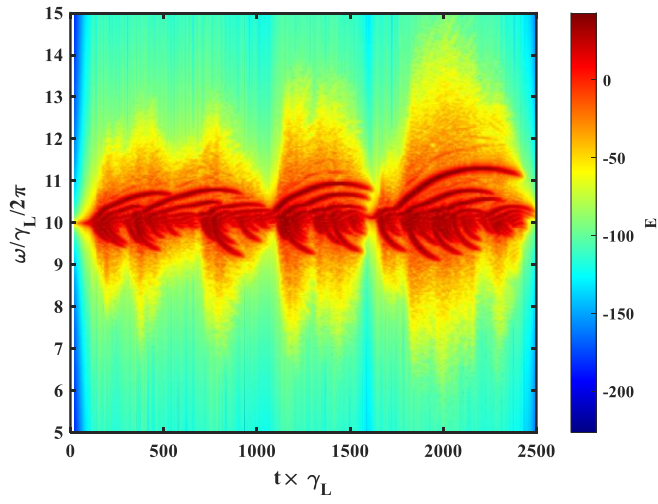

(b)

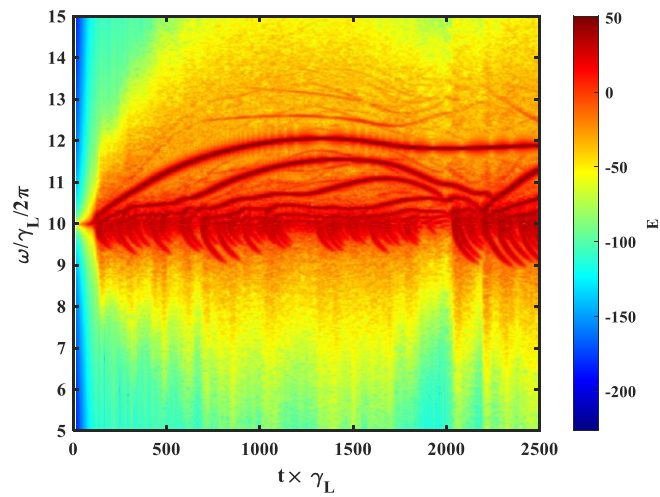

(c)

Figure 16. Spectrograms corresponding to the three cases in Fig. 15.

Beyond identifying chirping transitions at fixed diffusion strength, it is also important to quantify how the overall magnitude of collisions affects the saturation amplitude when the ratio $\alpha/\nu$ is held fixed. This provides a bridge from qualitative categorization to quantitative scaling. To keep the mode in a steady state while increasing collision strength, it becomes necessary to increase the diffusion operator (e.g., to 5 and 10), as shown in Fig. 17. The key observation is that, for the same $\alpha/\nu$ ratio, a larger diffusion operator leads to a larger wave amplitude. This trend can be understood through the restoration-versus-flattening balance: the collision operators represent the restoration rate of the distribution function, while the wave field represents the flattening rate due to resonant interaction. When collisions restore the gradient more rapidly, the wave must grow to a higher amplitude to maintain a balance that yields saturation, consistent with the interpretation of saturation physics discussed in Ref. [35].

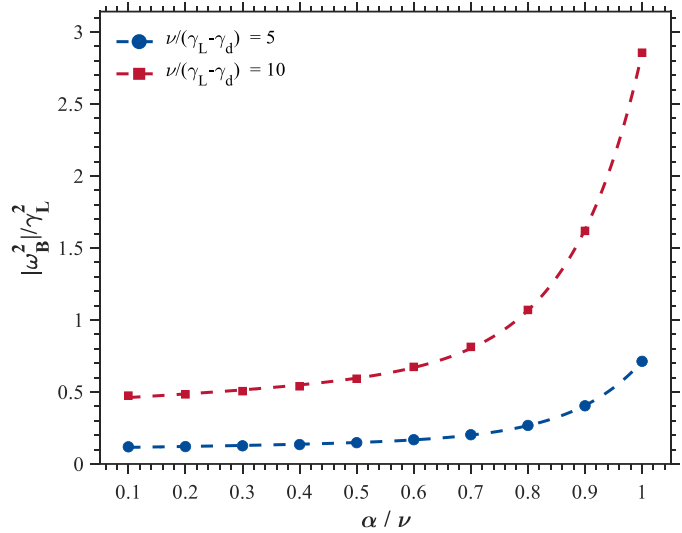

Figure 17. Steady-state saturation amplitude $C_{\text{sat}}$ versus the drag-to-diffusion ratio $\alpha/\nu$ at $\gamma_d/\gamma_L = 0.9$, shown for $\nu/(\gamma_L - \gamma_d) = 5$ and 10.

The corresponding regime distribution across parameter space is summarized in the bifurcation diagram shown in Fig. 18 (for the representative case with $\nu/(\gamma_L - \gamma_d) = 1$ held fixed), which consolidates the regime boundaries inferred from the time traces, spectra, and phase-space structures. As can be seen, lower damping generally raises the upper boundary for mode conversion, meaning the nonlinear response is more likely to remain bounded and effectively "more stable"; meanwhile, as $\alpha/\nu$ increases, the system follows an ordered transition from steady to periodic and then to chaotic behavior [16]. In addition, for $\gamma_d/\gamma_L \leq 0.3$, once chirping appears it tends to become persistent, corresponding to long-lived hole-clump pairs in phase space: under weak damping the wave energy is larger (as reflected by the trends shown in Figs. 7 and 12), and a stronger wave can modify the effective background dissipation, while the hole-clump structure carries negative wave energy and thus provides a feedback that further amplifies the hole-clump growth; moreover, when the wave field exhibits periodic behavior, its characteristic frequency also shifts, and one can observe a transition from intermittent chirping to persistent chirping. Unlike some previous findings, we also observe that, near marginality, the system can jump directly from a damped state to a chaotic state; we attribute this to the relatively small diffusive collisionality adopted here, since weaker relaxation at small scales makes the mode more prone to abrupt state switching and thus more likely to enter instability without passing through a gradual periodic stage.

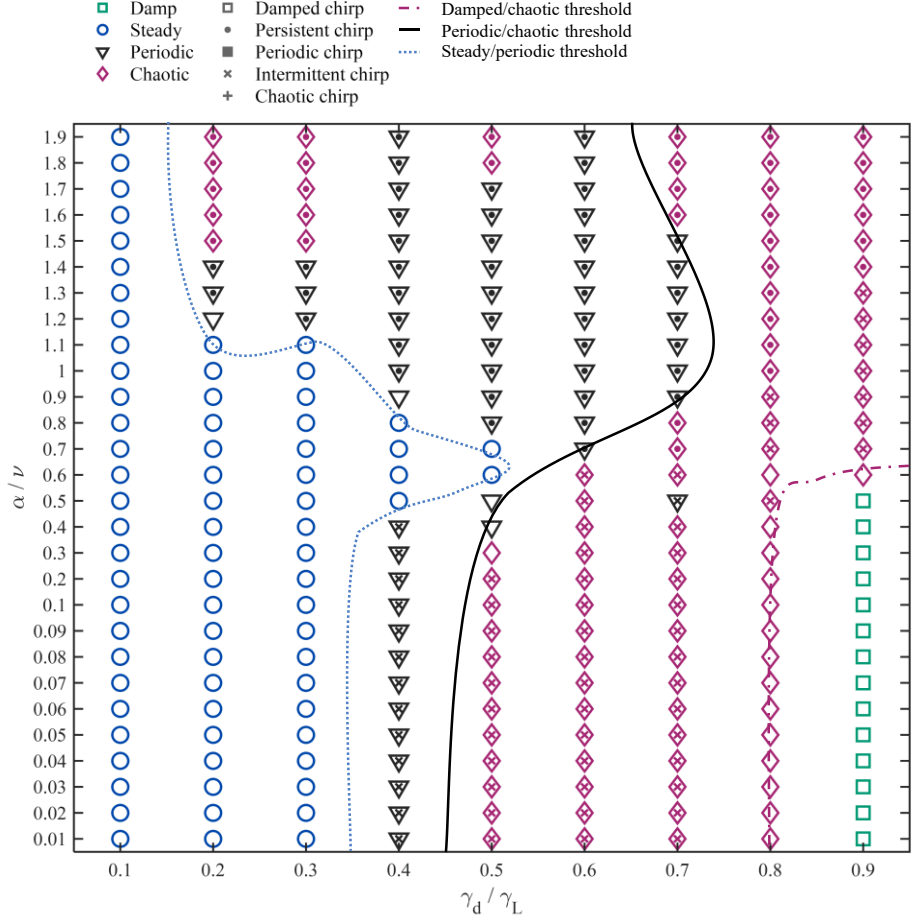

Figure 18. Behavior bifurcation diagram in the $(\gamma_d/\gamma_L, \alpha/\nu)$ parameter plane for the diffusion-drag model at fixed $\nu/(\gamma_L - \gamma_d) = 1$. The outer marker identifies the primary regime (damped, steady, periodic, chaotic), while the inner marker indicates the corresponding chirping subcategory (damped, periodic, intermittent, persistent, chaotic chirping), as defined in Sec. 3.1.

**3.3.2 Effects of Krook and drag collisions**

We next examine the combined effects of Krook and drag collisions. This case is particularly instructive because Krook relaxation and diffusion can both eliminate holes and clumps, yet they operate through different physical actions and timescales. Specifically, the diffusion timescale scales as $1/\nu^3$, whereas the Krook relaxation timescale scales as $1/\beta$. As a consequence, for comparable normalized operator values, the Krook term can act as a substantially more effective "restoring" mechanism than diffusion, thereby altering the survival time and spatial extent of phase-space structures and, in turn, the observed chirping behavior.

This difference is evident in Fig. 19, which shows the three-dimensional evolution of the distribution function near the threshold $\gamma_d/\gamma_L = 0.9$ under several representative combinations of $\alpha$ and $\beta$. In Fig. 19(a), a chaotic solution is obtained only when $\beta/(\gamma_L - \gamma_d)$ is sufficiently small; in that case, the distribution exhibits intermittent hole-clump structures and the system experiences episodes where perturbations appear to vanish. Notably, during the interval $\tau \times \gamma_L \approx 1850 - 2100$, an "empty window" forms in which the distribution returns close to equilibrium, followed by re-excitation of the instability and the onset of a new cycle. This behavior provides a clear phase-space interpretation of intermittent chirping: chirping activity is not continuously sustained but is repeatedly quenched by restoration and then reignited when sufficient drive is rebuilt. When the ratio $\alpha/\beta$ reaches unity (Fig. 19(b)), the region of velocity space affected by the perturbation is significantly smaller than in the diffusion-drag case of Fig. 15(b), and the lifetime of individual hole structures becomes shorter. Increasing $\alpha/\beta$ further to 15:1 (Fig. 19(c)) prevents the formation of a stable hole altogether. In terms of the secondary categorization introduced in Sec. 3.1, these phase-space changes correspond to a transition from intermittent chirping toward chaotic chirping, and they also indicate that sufficiently strong drag (relative to Krook) is required before the system can enter a persistent chirping regime. In other words, Krook relaxation tends to "reset" the phase-space perturbation before long-lived coherent structures can dominate, thereby shifting the regime boundaries compared with the diffusion–drag case.

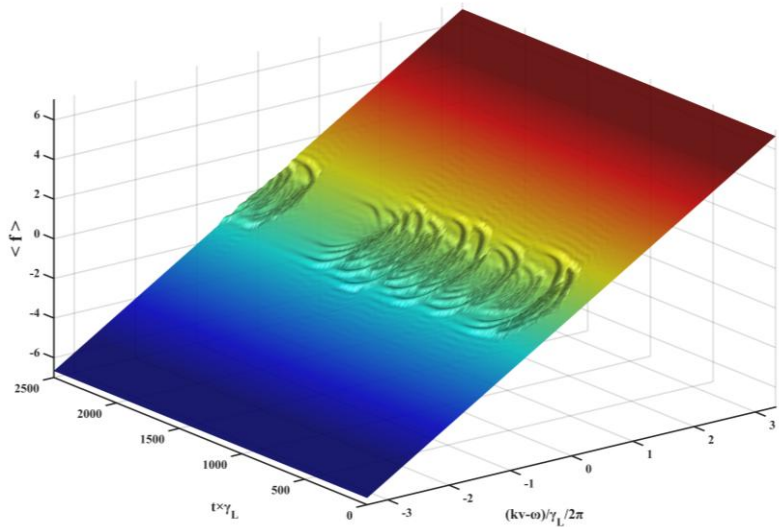

(a)

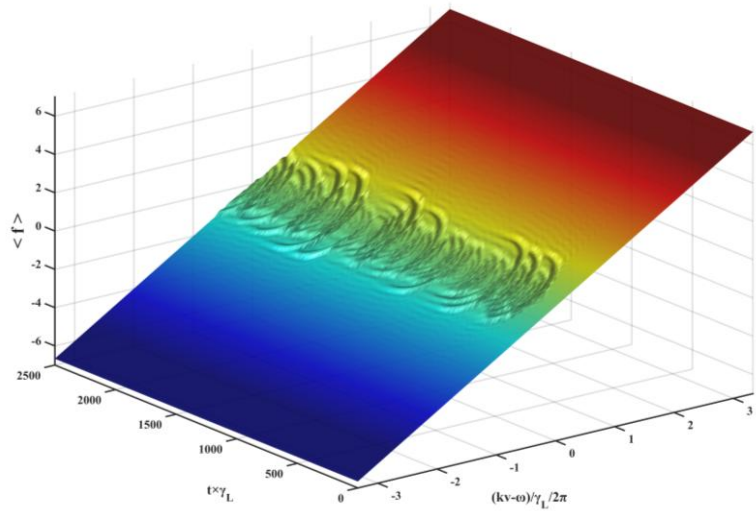

(b)

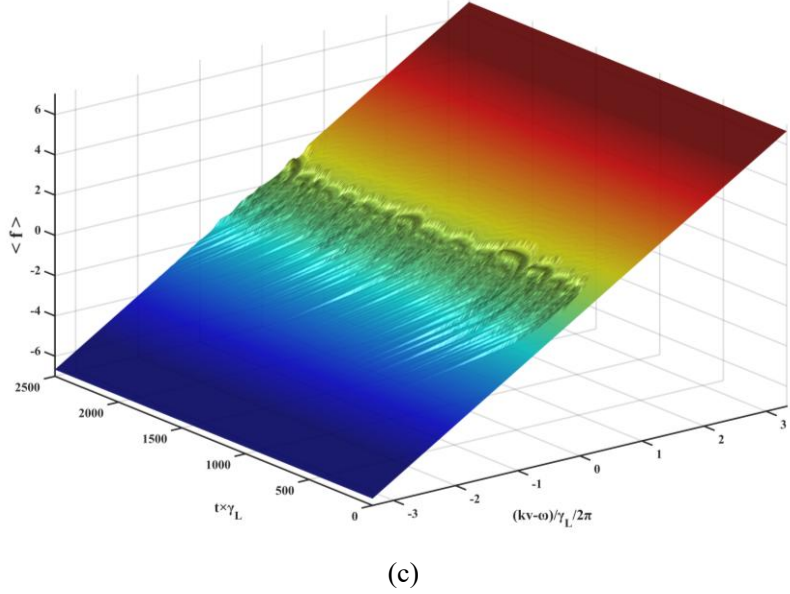

(c)

Figure 19. Three-dimensional evolution of the distribution function for the combined Krook–drag model at $\gamma_d/\gamma_L = 0.9$ and fixed $\beta/(\gamma_L - \gamma_d) = 0.1$: (a) $\alpha/(\gamma_L - \gamma_d) = 0.01$; (b) $\alpha/(\gamma_L - \gamma_d) = 0.1$; (c) $\alpha/(\gamma_L - \gamma_d) = 1.5$.

The corresponding spectral evolution is presented in Fig. 20. The chirping pattern progresses from a relatively symmetric structure (Fig. 20(a)) to an asymmetric state exhibiting both up-chirping and down-chirping (Fig. 20(b)), and finally to a regime where down-chirping disappears (Fig. 20(c)), consistent with the vanishing of the clump component. This spectral sequence provides a direct bridge between the phase-space interpretation (clump suppression and hole dominance) and the observed frequency dynamics. Importantly, it reinforces the general principle established in Sec. 3.2.3: drag promotes convective persistence and asymmetry, whereas a strong restoring operator (here Krook) suppresses long-lived clumps and can truncate chirping branches.

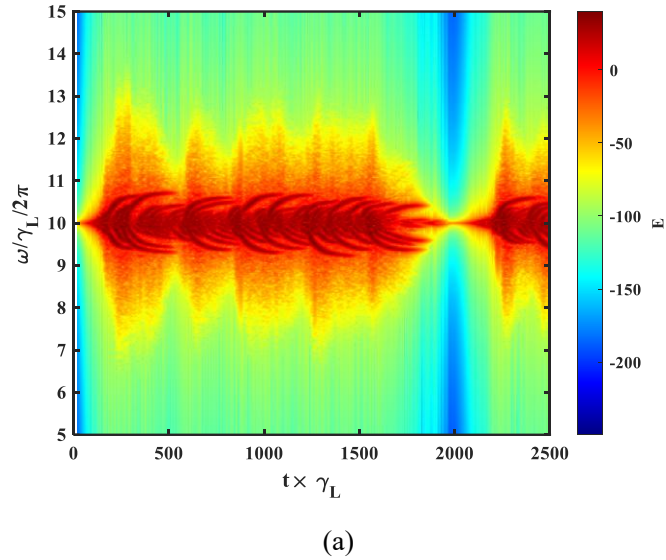

(a)

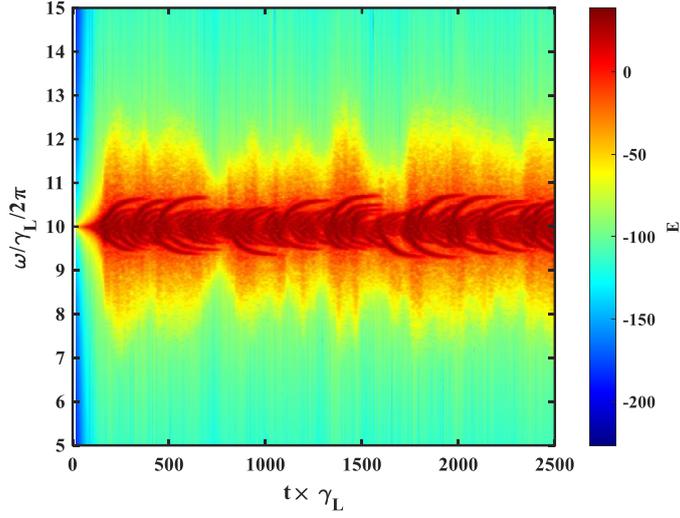

(b)

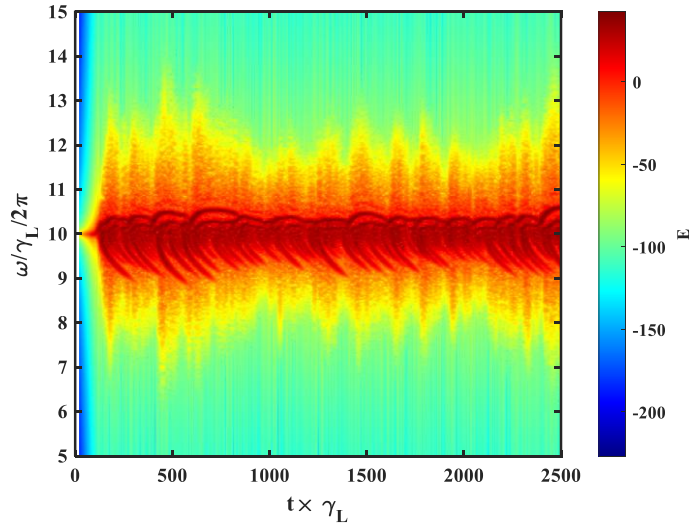

(c)

Figure 20. Spectrograms corresponding to Fig. 19 for the Krook–drag model

Finally, Fig. 21 complements the near-threshold structural analysis by quantifying how the wave amplitude varies across different $\alpha/\beta$ ratios. Similar to Fig. 17, an approximately exponential dependence of wave amplitude on the collision-operator ratio is observed; however, the physical interpretation differs because Krook and diffusion do not act in the same way. According to Eq. (2.4), diffusion acts on the gradient of the distribution function, while the Krook operator acts directly on the distribution function itself. Therefore, even under the same nominal restoration rate implied by an identical drag-to-operator ratio, diffusion tends to suppress nonlinear phase-space structures more strongly than Krook does. As a result, the drag-diffusion system typically reaches saturation (or collapse) earlier, whereas the drag-Krook system allows nonlinear structures to evolve to higher energy levels and thus supports larger wave amplitudes.

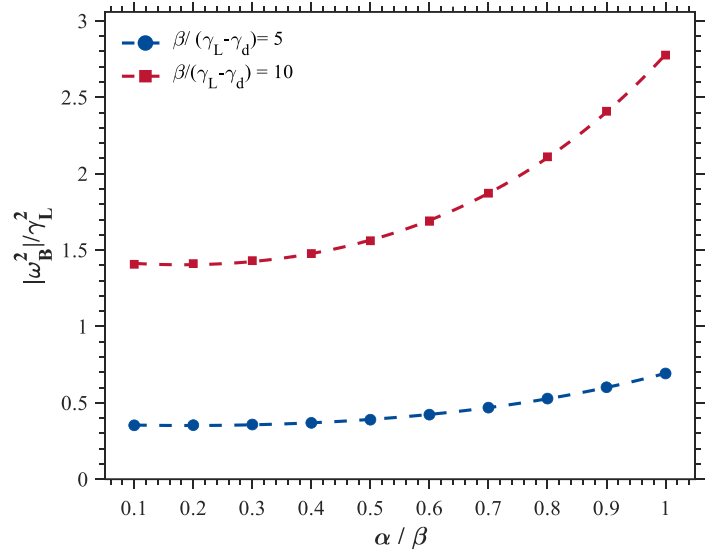

Figure 21. Steady-state saturation amplitude $C_{\text{sat}}$ as a function of the drag-to-Krook ratio $\alpha/\beta$ at $\gamma_d/\gamma_L = 0.9$, shown for $\beta/(\gamma_L - \gamma_d) = 5$ and 10.

    The overall regime distribution and transitions in the combined-operator parameter space are summarized in the behavior bifurcation diagram of Fig. 22, which consolidates the role of Krook restoration in shifting the boundaries among intermittent, chaotic, and persistent chirping regimes. In contrast to Fig. 18, pronounced frequency sweeping can be strongly suppressed over substantial portions of this scan because, for comparable normalized operator strengths, the Krook term acts as a substantially more effective restoring mechanism than diffusion, thereby shortening the survival time and spatial extent of hole-clump structures and repeatedly "resetting" the phase-space perturbation before long-lived coherent pairs can dominate. Nonetheless, Fig. 22 also exhibits a transition of the primary response from chaotic to periodic behavior, which is qualitatively consistent with the diffusion-drag map and can be interpreted as the outcome of competition between drag-induced convective deformation and Krook-driven relaxation: this competition accelerates energy transfer among the wave, energetic particles, and the background distribution and, as a direct consequence, confines the accessible wave-amplitude excursion, so that within the same observation time the chaotic field no longer wanders irregularly across a broad interval from $C_{\min}$ up to intermediate levels and ultimately toward $C_{max}$, but instead locks into a regular oscillation between two bounded levels $C_1$ and $C_2$, satisfying $C_{\min} < C_1 < C_2 < C_{max}$.

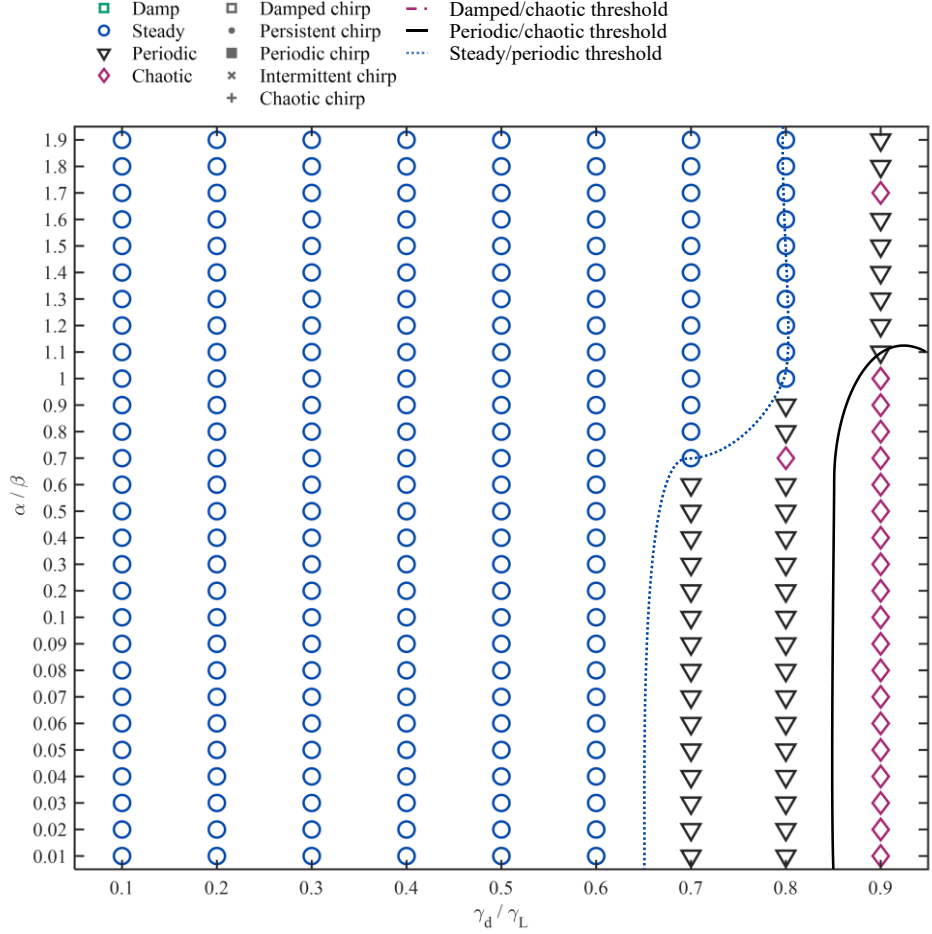

Figure 22. Behavior bifurcation diagram in the $(\gamma_d/\gamma_L, \alpha/\beta)$ parameter plane for the drag-Krook model at fixed $\beta/(\gamma_L - \gamma_d) = 1$.

### 3.3.3 Effects of diffusion, Krook and drag collisions

To assess the combined action of drag, diffusion, and Krook relaxation, we consider a representative case in which all three collision operators are retained with comparable characteristic timescales. We find that the resulting nonlinear dynamics do not introduce qualitatively new regimes beyond those identified in the pairwise studies. Instead, the primary effect of the three-operator interaction is a systematic shift of regime boundaries: diffusion and Krook relaxation partially counteract drag-induced convective deformation, increasing the wave amplitude required to sustain persistent chirping. This observation suggests that multi-operator effects can be interpreted as a competition between collision-driven restoration and drag-induced resonance distortion, rather than as a fundamentally new nonlinear mechanism. Figure 23 illustrates a representative transition sequence using a controlled scan where the operator ratios are held fixed while their absolute strengths are increased: for a baseline setting with comparable normalized magnitudes ($\alpha:\beta:\nu$ =1:1:1), the distribution function clearly forms a hole–clump pair and the wave field remains irregular with evident frequency excursion when $\alpha/(\gamma_L - \gamma_d) = 0.1$, consistent with intermittent chirping. Moreover, when all three operators are increased by a factor of ten at the same ratios, the hole-clump structures are suppressed and the dynamics regularize into a periodic regime; upon a further tenfold increase, a pronounced resonant plateau (distribution flattening) develops, the accessible drive is reduced, and the system approaches steady-state saturation.

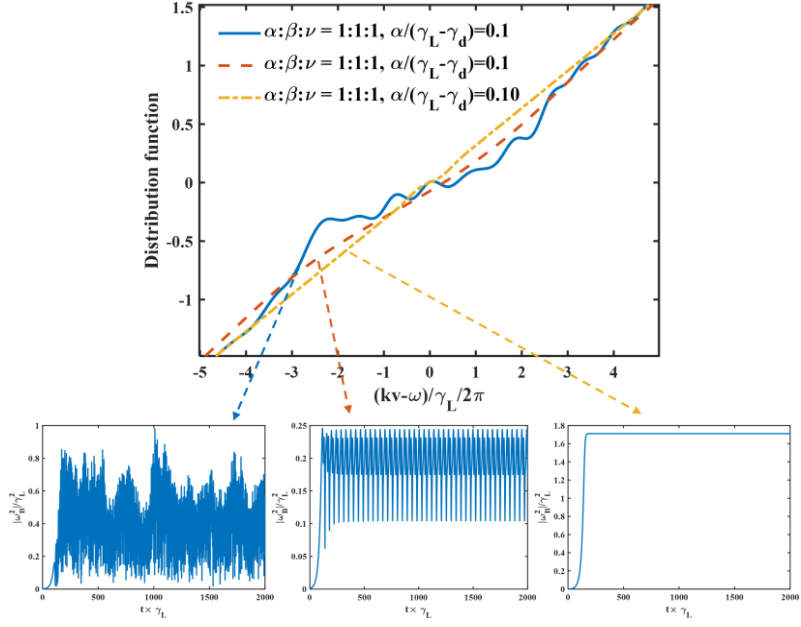

Figure 23. Uniform scaling of drag, Krook, and diffusion (α: β: ν = 1: 1: 1) induces a mode transition.

## 3.4 Experimental Verification and Perspectives

From an experimental perspective, these results suggest that increasing effective energetic-particle slowing-down (modelled here by drag) favours persistent chirping, whereas enhanced pitch-angle scattering or decorrelation mechanisms (modelled by diffusion or Krook relaxation) promote transitions toward periodic or steady behaviour. Such trends are consistent with experimental observations in which changes to auxiliary heating or background plasma conditions modify chirping activity. Experimental verification of the regime maps and chirping taxonomy developed in Secs. 3.1-3.3 can proceed at two complementary levels: phenomenological classification from diagnostics, and quantitative parameter inference. On the phenomenological side, experimental magnetic (Mirnov), soft-x-ray, reflectometry, or ECE signals can be transformed into spectrograms and classified using the same two-level framework adopted here: namely, the primary nonlinear regimes inferred from long-time wave-amplitude behavior and the secondary chirping categories inferred from the organization and persistence of frequency sweeping. This is directly aligned with the operational definitions introduced in Sec. 3.1, where chirping is treated as a subtype spanning multiple primary regimes rather than being restricted to a single nonlinear state. Such a unified classification enables a consistent comparison between numerical bifurcation diagrams and experimentally observed transitions among damped, steady, periodic, and chaotic responses and their corresponding chirping manifestations. On the quantitative side, several established "spectroscopic" approaches show that measured chirping characteristics can be inverted to estimate internal mode amplitudes and kinetic parameters that enter Berk-Breizman-type reduced models. For example, frequency sweeping in MAST has been interpreted within the Berk-Breizman paradigm, reproducing burst sequences and nonperturbative chirping features, thereby providing a natural benchmark for regime identification [7]. Likewise, persistent up-down chirping observed in JET has been attributed to the formation and evolution of resonant phase-space structures (holes and clumps) that continuously shift their characteristic frequency [39]. In addition, spectroscopic methods have been developed to infer absolute internal amplitudes from chirp rates, offering a direct bridge between external measurements and core fluctuation levels [25]. More recently, fitting strategies have been proposed to extract linear drive and external damping rates by matching observed chirping laws to Berk-Breizman simulations, and have been applied to chirping TAEs (e.g., in JT-60U) [40]. These tools suggest a concrete experimental validation path for our multi-operator predictions. In particular, controlled scans that vary effective collisionality and slowing-down (via density, impurity content, or beam and ICRF tailoring) and external damping (via equilibrium and continuum coupling) can be mapped onto the normalized parameters used here, enabling direct tests of predicted transition sequences. For instance, our simulations indicate that increasing drag at fixed diffusion can drive systematic changes from transient to sustained chirping, and can generate hook-like patterns that resemble experimentally reported structures. Beyond validation, main perspective is to generalize this framework to more realistic Alfvenic systems: electromagnetic fields, toroidal geometry, multiple resonances, and multi-mode coupling. The same classification and inference workflow can then be integrated with interpretive modeling to diagnose when chirping implies long-lived phase-space structures and when it reflects intermittent, near-threshold relaxation, ultimately improving predictive capability for energetic-particle transport in reactor-relevant regimes [41].

## 4 Conclusion

This work aimed to close a key gap in the nonlinear description of the Berk-Breizman BOT instability by establishing a unified and operational classification, and then using it to determine how external damping and collisional relaxation, acting individually and in combination, govern saturation, chirping, and regime transitions. Building on the fully kinetic wave model and the numerical scheme described in Sec. 2, we introduced a two-level categorization that (i) assigns primary regimes (damped, steady state, periodic, chaotic) from the long-time evolution of the normalized wave energy and (ii) resolves chirping as a secondary subtype spanning multiple primary regimes (damped, persistent, periodic, intermittent, chaotic chirping) based on spectrogram organization and the persistence of frequency shifts. Importantly, this two-level categorization separates the global energy balance outcome from the underlying resonance dynamics, so chirping is treated as a nonlinear manifestation that can arise across regimes rather than being a regime by itself. Using this framework, we demonstrated that diffusion and Krook relaxation both promote regularization and steady saturation. We interpret both operators as collision driven restoration mechanisms, while noting that they implement restoration through different phase space pathways. Diffusion preferentially smooths fine gradients and suppresses long lived coherent structures, whereas Krook relaxation restores the distribution more uniformly and allows larger perturbation levels before a steady balance is reached. In both cases, the steady saturation amplitude exhibits a strong and approximately exponential dependence on $\gamma_d/\gamma_L$. In sharp contrast, drag acts convectively in velocity space and produces pronounced convective deformation of resonant phase space structures. This convective deformation intrinsically sustains strong time dependence, eliminates steady state solutions in the scanned parameter ranges, and produces chirping dominated chaotic or explosive behavior together with substantial deformation of the resonant region. When multiple collision operators are present, we showed that nonlinear behavior is not a linear superposition of single operator limits. Drag-diffusion interactions systematically break hole-clump symmetry and yield ordered chirping transitions with increasing $\alpha/\nu$, reflecting how diffusion driven restoration competes with drag driven convective deformation. Drag-Krook interactions favor intermittent re-excitation and support larger amplitudes than drag-diffusion at comparable operator ratios, consistent with the different restoration mechanisms and their associated timescales. The resulting bifurcation diagrams provide a predictive and quantitative atlas of nonlinear regimes in collision parameter space and clarify the controlling physical principle. The observed state is set by the competition between wave driven resonance flattening and collision driven restoration, and this balance is further reorganized by drag induced convective deformation. In summary, this work provides a unified nonlinear framework for interpreting energetic-particle-driven chirping within BOT paradigm when multiple collision mechanisms act simultaneously. By separating global nonlinear regimes from resonance-scale chirping dynamics and systematically mapping collision-operator effects, we clarify why drag promotes persistent chirping while diffusion and Krook relaxation favour saturation. These results offer predictive guidance for interpreting and potentially controlling chirping Alfvénic activity in magnetically confined fusion plasmas. Future work should extend these regime maps to electromagnetic energetic particle modes, incorporate more realistic resonance structures including resonance line splitting and friction effects, and couple to global or gyrokinetic simulations and experimental diagnostics to translate the present phase space categorization into predictive control strategies for chirping and energetic particle transport.

## Acknowledgement

Authors appreciate Matthew Lilley for providing BOT code and inspiring discussions during Lei Chang's visit in UT-Austin. The work was supported by National Natural Science Foundation of China (Grants No. 92271113, 12411540222, and 12481540165), Natural Science Foundation Project of Chongqing (Grant No. CSTB2025NSCQ-GPX0725), and ENN Hydrogen–Boron Fusion Research Fund (Grant No. 2025ENNHB01-011).

## Data availability statement

The data that support the findings of this study are available from the corresponding author upon reasonable request. Simulation codes used in this work are based on the publicly available BOT code developed by Matthew Lilley and collaborators [17,23,34-38], which can be accessed at https://github.com/mklilley/BOT